**Refractive uses of layered and two-dimensional materials for integrated photonics**

Akshay Singh[1,*], Seong Soon Jo[2], Yifei Li[2], Changming Wu,[3] Mo Li,[3,4] R. Jaramillo[2,*]

1. Department of Physics, Indian Institute of Science, Bengaluru, India

2. Department of Materials Science and Engineering, Massachusetts Institute of Technology, Cambridge, MA, USA

3. Department of Electrical and Computer Engineering, University of Washington, Seattle, WA, USA

4. Department of Physics, University of Washington, Seattle, WA, USA

*Corresponding Authors

**Abstract:** The scientific community has witnessed tremendous expansion of research on layered (*i.e.* two-dimensional, 2D) materials, with increasing recent focus on applications to photonics. Layered materials are particularly exciting for manipulating light in the confined geometry of photonic integrated circuits, where key material properties include strong and controllable light-matter interaction, and limited optical loss. Layered materials feature tunable optical properties, phases that are promising for electro-optics, and a panoply of polymorphs that suggest a rich design space for highly-nonperturbative photonic integrated devices based on phase-change functionality. All of these features are manifest in materials with band gap above the photonics-relevant near-infrared (NIR) spectral band ($\sim 0.5 - 1$ eV), meaning that they can be harnessed in refractive (*i.e.* non-absorptive) applications.

# Introduction: Materials needs for photonics

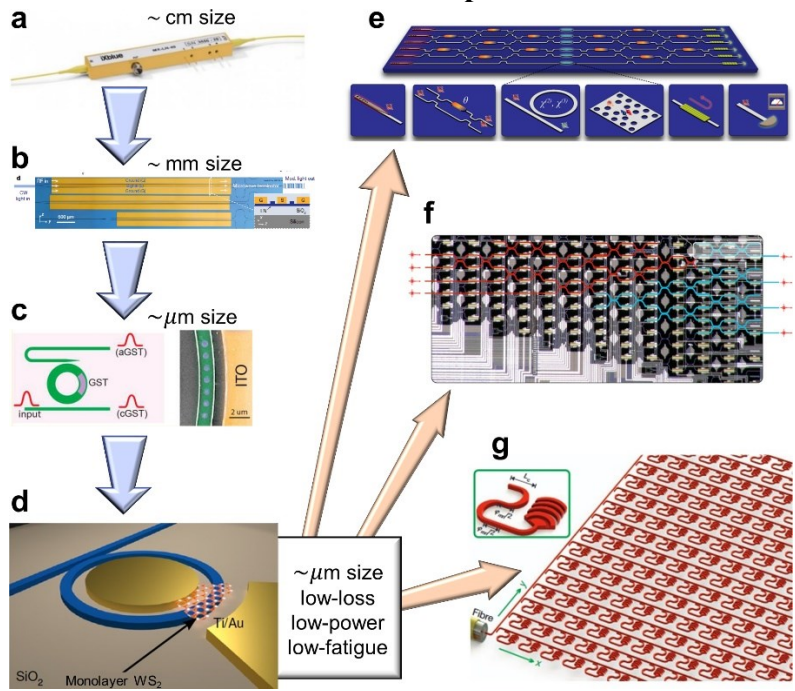

**Figure 1:** Materials needs for integrated photonics. (a-d) Making compact, low-loss optical phase modulators through materials innovation. (a) Commercial modulator based on LiNbO$_3$. (b) Phase modulator with reduced footprint, realized by integrated processing on LiNbO$_3$ wafer; reprinted with permission from Springer Nature.[1] (c) Integrated photonic switch enabled by phase-change materials; reprinted from *ACS Photonics*.[2] (d) Future materials will enable integrated optical phase modulators with fast switching speed, low insertion loss, low power consumption, low fatigue, and excellent process compatibility; reprinted with permission from Springer Nature.[3] (e-g) Future technologies enabled by materials innovations. (e) Quantum photonic integrated circuits; reprinted with permission from The Optical Society.[4] (f) Neuromorphic processors; reprinted with permission from Springer Nature.[5] (g) Beam steering for navigation and telecommunications; reprinted with permission from Springer Nature.[6]

Photonic integrated circuit technology has achieved a remarkable level of maturity. Recent highlights include quantum emitters, neuromorphic computing accelerators, and phased arrays for beam steering.[4–6] However, in key respects we remain at an early stage of development. In particular, we have not yet developed materials to enable strong optical phase modulation with the speed, insertion loss, power consumption, endurance, and process compatibility needed for future, fully-integrated photonic circuits, processors, and sensors. For present-day mature photonics applications, devices such as switchable mirrors often remain discrete components. It is as one might imagine electronic integrated circuits without integrated diodes and transistors, with these essential devices remaining as discrete elements soldered to a circuit board. The analogy isn't perfect, but it does well to illustrate the urgent needs for better active materials for optical phase modulation in photonic integrated devices, and the tremendous upside potential of developing better materials-based solutions.

The requirements of strong optical phase modulation, fast and low-power switching, low material fatigue and long device lifetime, low insertion loss, and process compatibility result in a daunting set of technical specifications for integrated photonics active materials. Present-day solutions often satisfy some of these requirements, but not all. Silicon-based electro-optic phase shifters can be fast and highly-reliable, but are also power-hungry and the underlying effects are weak.[7,8] Optical phase shifters based on the Pockels effect in lithium niobate (LiNbO$_3$) are fast and highly-reliable, with low insertion loss, but the weak modulation requires large interaction length, and the material is notoriously difficult to integrate.[9] Optical modulators based on tunable plasmonic resonances can be fast and operate at low power, although optical loss remains a challenge.[10–14] Optical phase shifters based on thermo-optic effects can be compact and reliable, with good modulation strength and low insertion loss, but are inherently slow and power-hungry.[5,15] Phase-change functionality in materials such as Ge$_2$Sb$_2$Te$_5$ (in the Ge-Sb-Te (GST) system) enables strong optical phase modulation, and the materials are amenable to process integration and device size scaling, but suffers from high insertion loss and materials fatigue.[16,17] Clever photonic device engineering can mitigate these downsides, but cannot resolve the underlying materials drawbacks.[2]

Here we review the prospects for materials with layered crystal structures as active, refractive materials for integrated photonics. Layered materials (LMs) often feature remarkable optical absorption, dominated by excitonic properties, for which research and applications have been reviewed elsewhere.[18,19] Motivated by the need for devices with low insertion loss, we focus instead on the lesser-studied refractive properties, and light-matter interactions with photon energy below the band gap. This results in an emphasis on the NIR optical properties of materials with band gap above 1.5 eV. Layered chalcogenides have a large real part of refractive index ($n$), which is needed for mode confinement and strong interactions in the confined geometry of integrated photonic waveguides and devices. Recent results illustrate that external perturbation can produce substantial refractive index and optical phase shifting without appreciable optical loss: large $\Delta n$ without large $k$ (imaginary part of refractive index).[3] Work on phase-change properties of transition metal dichalcogenides (TMDs) holds promise for highly-non-perturbative optical modulators.[20–22] LMs also have intriguing electro-optic properties as well as phase-sensitive optical components, which we also review here.

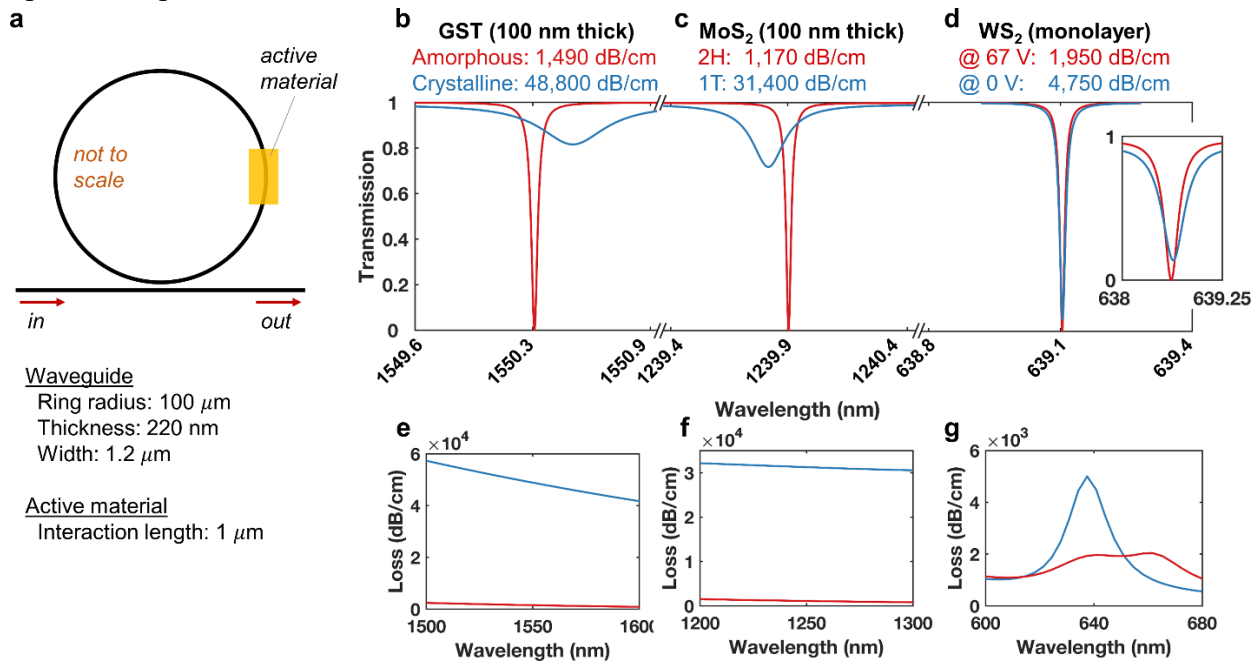

**Figure 2:** Role of materials offering strong optical phase modulation for photonic integrated devices, illustrated by simulated ring-resonator devices. (a) Schematic and key dimensions of simulated devices, showing location of the patch of active material that is laid on top of the $Si_3N_4$ waveguide. (b-d) Transmission spectra for active materials $Ge_2Sb_2Te_5$, $MoS_2$, and $WS_2$; see text for details. The axes titles report loss figures on-resonance. (d, inset) Transmission spectra for $WS_2$ with interaction length increased to 8 $\mu$m. (e-g) Loss spectra corresponding to panels (b-d). The loss data report loss (in dB) per length of active material (in cm).

In **Fig. 2** we illustrate the usefulness of materials that offer strong optical phase modulation for applications in photonic integrated devices. We use Lumerical software to simulate results for a prototypical $Si_3N_4$ ring resonator with a covering patch of active material, as illustrated in **Fig. 2a**. We consider three active materials, using $n, k$ data available in the literature: 10 nm-thick $Ge_2Sb_2Te_5$, 10 nm-thick $MoS_2$, and monolayer $WS_2$.[23–25] For GST and $MoS_2$ we simulate a wavelength range in the NIR, below the band gap in the low-loss phase of each material. For $WS_2$ we simulate results for visible light, at the tunable exciton absorption resonance. GST offers strong optical phase modulation when changing between amorphous and crystalline phases, due to the large $\Delta n$, and resulting in a substantial shift in the resonance position. However, the crystalline phase is optically lossy (large $k$), resulting in a broad resonance and substantial insertion loss. $MoS_2$ also offers strong optical phase modulation when it changes between crystalline phases (*i.e.* the 2H and 1T polymorphs). Compared to GST, $MoS_2$ has reduced loss and a sharper resonance in the lossy phase (crystalline for GST, 1T for $MoS_2$). Monolayer $WS_2$ has a much weaker effect on the simulated device performance, due to the small interaction volume. The resonance is sharp, but barely shifts even for a very large electrostatic gate bias. If the interaction volume is increased with a longer interaction length, then the resonance becomes less sharp due to the strong absorption at the exciton resonance (**Fig. 2d**, inset). These simulations - and recent experimental results on $WS_2$ optical phase shifters (see **Fig. 10** and ref. 3) - motivate continued research on refractive applications of TMDs for photonic integrated devices, and the focus on phase-change functionality.

$k$ is a critical parameter for refractive materials for integrated photonics, but is difficult to measure in the below-band gap spectral range. Experimental measurements usually depend on discerning a small signal from a large background. Furthermore, $k$ below-band gap is highly-dependent on materials processing (much more so than $n$), and therefore may have a large range of reported values for a given material. This is appreciated in traditional photonics, for which materials processing for low loss is essential; examples include MultiSpectral (TM) ZnS ceramics, processed for low-loss by reducing sulfur vacancy concentration, and chalcogenide glass fibers processed for low-loss by reducing oxygen contamination.[26,27] Similar efforts will be required to limit loss in below-band gap, refractive applications of LMs, but the materials science is at an early stage. We discuss techniques for measuring the optical constants and anisotropy of LMs, essential for materials qualification and selection. We discuss pros and cons of each technique, and motivate use of certain lesser-used but effective techniques.

We also devote attention to materials processing, with an emphasis on large-area coating methods appropriate for planar device fabrication, because even a spectacular material isn't viable for commercialization without process-compatibility. This review primarily covers monochalcogenides and dichalcogenides, with chemical formulae MX and $MX_2$, respectively (M-metal, X- chalcogen); we also discuss black phosphorus (bP), due to structural and functional similarities with some monochalcogenides.

We close this introduction with a brief discussion of the word "phase". In optics, the phase of light describes a relative position in a cyclic waveform. In materials science, the phase of a material is defined by its composition and structure, and determines its properties. Optical phase modulators change the phase of light relative to a reference beam. However, "phase change" may also imply functionality deriving from a materials phase transformation, such as crystalline-amorphous transformations in GST, or martensitic transformations in TMDs. In an attempt to reduce confusion, throughout this review, we use the term "optical phase" to refer to the phase of light, and "phase-change" everywhere we refer to a materials phase transformation.

**Optical properties of layered materials**

*Introduction to layered materials*

LMs are studied widely due to their diverse physical properties, which become particularly rich in the two-dimensional (2D) limit of individual monolayers. LMs typically have weak van der Waals (vdW) bonding between layers, and strong bonding within layers, resulting in highly-anisotropic optical and electronic properties. In **Fig. 3** we illustrate representative crystal structures for the LMs discussed here: dichalcogenides ($MX_2$), monochalcogenides (MX), and bP. All LMs are strongly biaxial (*i.e.* birefringent) due to their layered structure. Many also feature in-plane anisotropy, making them triaxial. LMs can often be prepared with varying layer count, all the way down to individual monolayers. The electronic structure can vary strongly with layer count; a well-known example is that of 2H-$MoS_2$, which as a bulk crystal has an indirect band gap with energy ($E_g$) of 1.3 eV, but as a 2D monolayer has a direct band gap of $E_g$= 1.8 eV.[28] LMs usually feature strong exciton absorption resonances and large exciton binding energy, which result from the weak dielectric screening for out-of-plane electric fields, which in turn follows from layered crystal structure. LMs also feature notable mechanical properties such as high flexibility and facile inter-layer sliding; $MoS_2$ is widely used as a dry lubricant (*e.g.* trademarks MOLY-MIST and Never-Seez Blue Moly). Their structure and opto-electronic properties can be controlled by chemical doping, surface modification, the underlying substrate, and by stimuli such as electric fields and mechanical strain.[18] Many LMs have $E_g$ in the range of 1 - 2 eV, making them attractive for refractive NIR applications.[29,30]

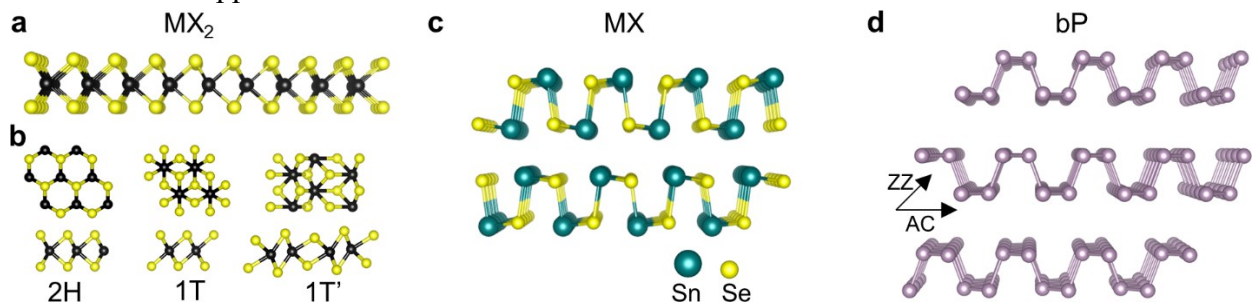

**Figure 3:** Representative crystal structures of the LMs discussed in this review: LMs with band gap in the NIR-VIS and strong light-matter interaction. (a) Side view of a monolayer in the trigonal prismatic (so-called 2H) phase common to many TMDs with stoichiometry *$MX_2$*. (b) Top-views (top row) and side views (bottom row) of common TMD polymorphs: 2H, 1T (tetragonal coordination), and 1T' (distorted 1T). (c) Monochalcogenides *MX* generally form as puckered layers with substantial in-plane anisotropy; here we show two layers of SnSe. (d) bP also adopts a puckered structure with high in-plane anisotropy; ziz-zag (ZZ) and armchair (AC) directions are indicated. The structure diagrams were generated using VESTA 3 software.[31]

| *Materials* | *Crystal structures* | *Anisotropy* | *Refractive index (n), in-plane (IP) and out-of-plane (OP)* |
|---|---|---|---|
| $MoS_2$, $MoSe_2$, $WSe_2$, $WS_2$ | $P6_3/mmc$; known as 2H; trigonal, prismatic coordination | Biaxial | 2-4 (IP)[24], 1-2 (OP, calculated) |
| $ZrS_2$, $ZrSe_2$; $TiS_2$, $TiSe_2$ | P-3m1; known as 1T, or the $Cd(OH)_2$ structure type; tetragonal coordination | Biaxial | 3~3.5 (IP)[32]; $TiS_2$ 3.9~4.9[24] |
| GeS, GeSe; SnS, SnSe | Pnma | Triaxial | GeS 3.3~3.9[33]; SnS 3.7-4.6[34] |
| bP | Cmce | Triaxial | bP 2~4[35] |

**Table 1:** Crystal structure, anisotropy and linear optical properties for materials discussed in this review. The crystal structure is related to the anisotropy, which relates to the different optical properties for in-plane (IP) and out-of-plane (OP) directions.

LMs feature band gap ($E_g$) ranging from 0 eV (metallic or semi-metallic, such as graphene and $TiS_2$) to 6 eV (hexagonal BN, hBN). We illustrate this distribution in **Fig. 4**. For some materials, we also illustrate the range of $E_g$ achievable by varying the layer count. Several trends are apparent in the data. For a given metal element, $E_g$ increases as the chalcogen varies from Te, to Se, to S. The band gap often changes from indirect to direct with decreasing layer count. Chemical alloying can tune $E_g$, traversing the difference between pure phases; examples include the $HfSe_2$-$HfS_2$, the $SnS_2$-$SnSe_2$, and the $ZrS_2$-$ZrSe_2$ systems.[32,36,37]

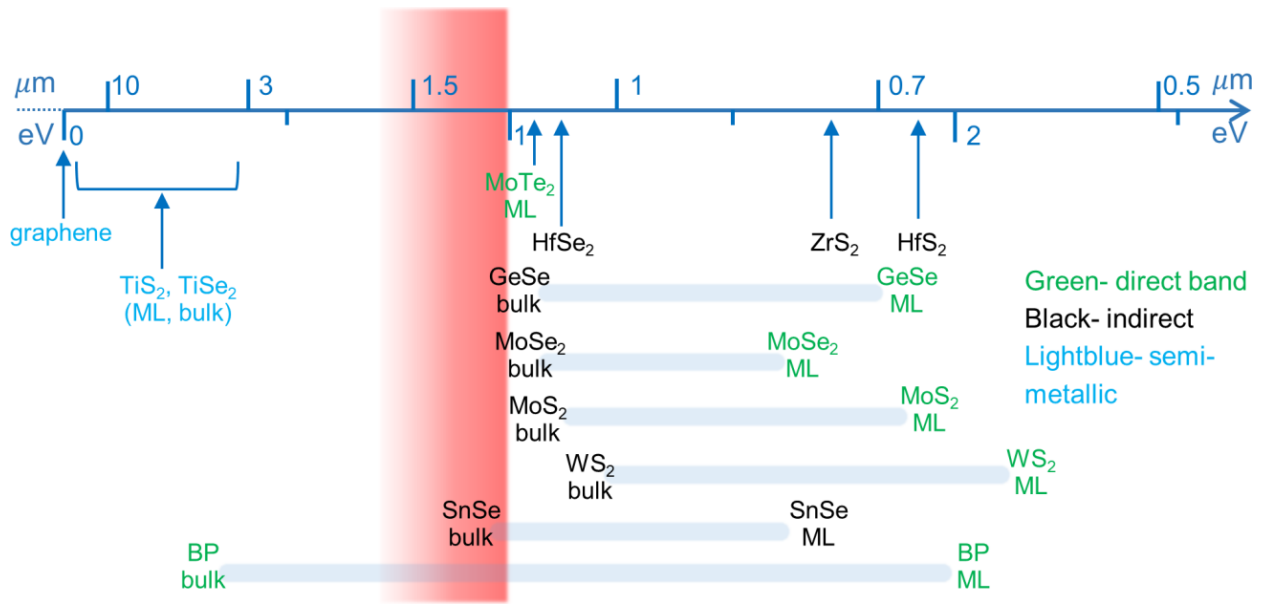

**Figure 4:** Distribution of experimentally measured band gap ($E_g$) for LMs discussed here. For some materials, we also present the variation of $E_g$ with layer count between the bulk and monolayer limits. For some materials the data reported are the exciton absorption resonances, because the fundamental band gap is difficult to determine experimentally. The labels are colored according to the nature of the gap: green = direct, black = indirect, light blue = metallic/semi-metallic. We also illustrate (in red) the NIR spectral range 0.75 – 1 eV relevant to telecommunications. Further details and data citations are presented in **Table 1**.

*Complex refractive index of layered materials*

The complex refractive index ($n - ik$) is related to the dielectric permittivity $\epsilon$ by $n = Re(\sqrt{\epsilon})$ and $k = -Im(\sqrt{\epsilon})$, and $k$ is related to absorption coefficient $\alpha$ by $\alpha = 4\pi k/\lambda$. For integrated photonics, materials with large $n$ are preferred to improve optical confinement and reduce mode leakage. There exists an extensive body of work measuring $n$ for LMs, dating back decades, focused mainly on spectral reflectivity and transmission in the visible range from basal faces of bulk LM crystals.[32,38,39] Measurements of ($n, k$) in the visible and NIR of few- and monolayer samples have appeared more recently, and reveal notable differences from bulk properties.[24,40–42] Optical properties can also be tuned with alloying, which greatly expands the materials design space. In **Fig. 5** we show two examples of optical properties tuning with alloying.[36,37]

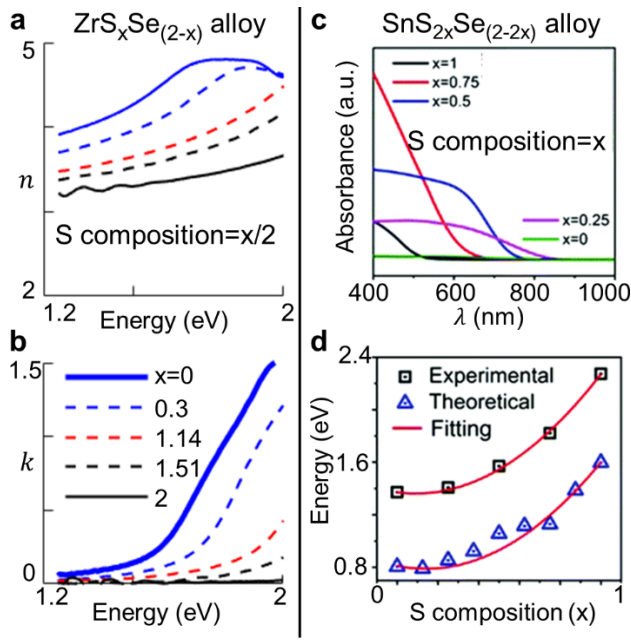

**Figure 5**: Alloying LMs provides remarkable control over sub-band gap optical properties. (a-b) Variation of $n, k$ with alloying in ZrS$_x$Se$_{2-x}$ alloys.[36] $n$ and $k$ both reduce with increasing sulfur content ($x$) in the NIR, due to the blue-shift of the exciton absorption resonance. (c-d) Variation of absorption spectra (c) and band gap (d) of SnS$_{2x}$Se$_{2(1-x)}$ alloys; reprinted with permission from the Royal Society of Chemistry.[37]

Here we report selected values of $k$ for LMs, but we caution the reader that these are contingent on measurement artifacts, and depend strongly on materials processing.[24,39] In our own research, we have found that value of $k$ that we determine for MoS$_2$ in the NIR band varies between 0.05 and 0.2, depending on materials processing and measurement technique (see section 2.6 for discussion of measurement techniques); the theoretically-predicted value is 0.01-0.02, suggesting that advances in materials processing will yield further improvements.[24] Going forward, a general rule for selecting new materials with acceptably-low $k$ for optical phase modulators for NIR integrated photonics is that the energetically-lowest absorption resonance (exciton or band gap) should be higher than 1.5 eV.

*Excitons and electron-hole screening*

Excitons are correlated electron-hole pairs bound by the Coulomb interaction. The electron-hole Coulomb interaction is screened by the dielectric constant of the material, and in most semiconductors the exciton binding energy is less than 0.1% of a Rydberg. The layered structure and vdW bonding of LMs decreases the screening effect and confines the electron and hole wavefunctions (see **Fig. 6a**), resulting in intra-layer excitons with large binding energy, even up to 1 eV. [43,44] These effects are most pronounced in 2D monolayers, but remain significant for bulk crystals. The large binding energy means that excitons in LMs tend to be stable at room temperature and often dominate the optical spectra. Excitons account for optical absorption resonances at energy ($E_X$) below the band gap ($E_g$), and strongly-enhanced light-matter interactions; the absorption coefficient ($\alpha$) on-resonance can exceed $10^5$ cm$^{-1}$.[28,41]

For refractive applications, strong exciton resonances can be a significant source of optical loss. The large exciton binding energy means that optical loss can be substantial even for photon energy well below $E_g$ and a defect-free material. This is illustrated by the theoretically-predicted optical properties of SnSe, calculated with and without exciton effects, presented in **Fig. 6b**.[30] The theoretical data show a substantial red-shift of the loss spectra, with clear implications for low-loss NIR photonics. However, the experimental challenge of measuring $E_g$ in the presence of strong exciton effects means that values reported as $E_g$ are often in fact $E_X$. As usual, in selecting materials it is advised to rely on the best available measurements.

Modulating exciton resonances can have a strong effect on the refractive index. Yu *et al.* reported that the refractive index of monolayer TMDs MoS$_2$, WS$_2$ and WSe$_2$ can be strongly modulated by electrostatic gating, with $\Delta n \sim \Delta k \sim 1$ in the vicinity of exciton absorption resonances (**Fig. 6c-e**).[25] This large tunability was attributed to broadening the linewidth of excitonic interband transitions due to neutral-charged exciton interconversion. Klein *et al.* demonstrated ultra-fast control of surface plasmon polariton (SPP) transmission on a gold waveguide by optically pumping the exciton resonance of an adjacent WS$_2$ monolayer.[10] Additionally, Liu *et al.* demonstrated field-effect tuning of a polariton resonance in a hybrid WS$_2$-plasmonic lattice.[11]

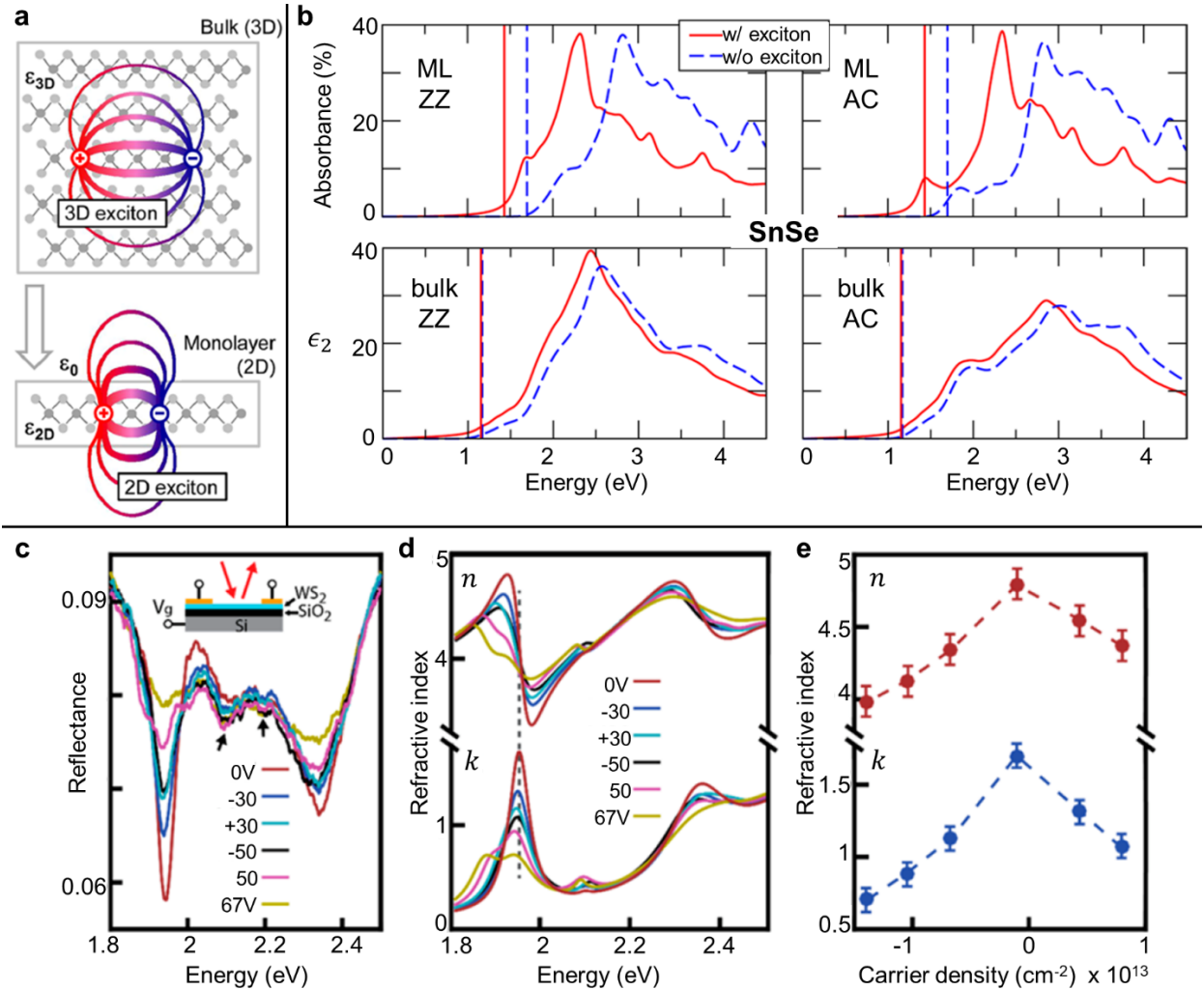

**Figure 6**: Exciton resonances have large effects on optical absorption, and must be treated properly when calculating the optical properties of layered materials. (a) Schematic showing the reduced dielectric screening in monolayers compared to bulk crystals; reprinted with permission from the American Physical Society.[43] (b) Calculated optical properties of SnSe with and without excitons; reprinted from *Nano Letters*.[30] The absorption onset is red-shifted for both bulk and monolayer materials. The effect of in-plane crystal anisotropy is apparent in the different optical spectra for electric field polarized along the zigzag (ZZ) and armchair (AC) directions. Field-effect tuning in the visible range (c-e); reprinted from *Nano Letters*.[25] (c) Reflectance change with gate voltage (line series); inset shows measurement geometry. (d) Change in the complex refractive index with gate voltage, illustrating the close correspondence between electro-refraction and electro-absorption near an exciton absorption resonance. (e) Dependence of the complex refractive index at-resonance on carrier density.

LMs can also be vertically integrated by mechanical stacking to create heterostructures. Such stacking can shift the band-gap due to changes in dielectric environment, and modify optical loss in NIR.[45] Sub-bandgap absorption can also result from inter-layer excitons, where electron and holes are separated and delocalized over two or more layers. These inter-layer excitons usually have low absorption and low binding energies, and therefore have lesser effects on optical loss, but could be a source of long-lived states which can interfere with fast modulation methods.[46,47] Dicke superradiance effects can also modify optical absorption in few layer TMDs; these effects are relatively well understood.[48]

*Bi- and tri-refringence*

The layered crystal structure of LMs leads to large birefringence. In **Fig. 7** we present experimental data for three LMs: hBN, SnS and bP.[34,49,50] For hBN, the birefringence in the VIS-UV has been measured to be $\Delta n = n_o - n_e \lesssim 1$ ($n_o$ and $n_e$ are the ordinary and extraordinary indices, respectively).[49] The band gap of hBN is very large ($E_g$= 6 eV, indirect), meaning that optical loss is low in the visible, and the large $\Delta n$ can be used for refractive applications. For SnS, $\Delta n \approx 0.5$ has been measured in the NIR. SnS has a rather low band gap ($E_g$ = 1.3 eV, indirect), but due to the indirect nature of the gap, the optical loss remains low for photon energy below 1 eV.[34] Experimental measurements of birefringence of TMDs are scarce (*infra vide*), but our theoretical calculations predict that $\Delta n \approx$ 1.7, 2.1, and 1.2 in the NIR for $MoS_2$, $TiS_2$, and $ZrS_2$, respectively.[24]

Many LMs have low in-plane symmetry due to ferroelastic distortions, making them triaxial. Examples include SnSe, SnS, GaS, and bP. The triaxial nature of SnS is apparent in the $(n, k)$ data presented in **Fig. 7b**. Mechanical strain can also be used to lower in-plane symmetry and induce triaxial properties in LMs.[51] Bi- and tri-refringence can be exploited for polarization- and phase-sensitive photonic components. **Fig. 7c** shows an example of in-plane anisotropy in bP exploited to generate elliptical polarization.[50]

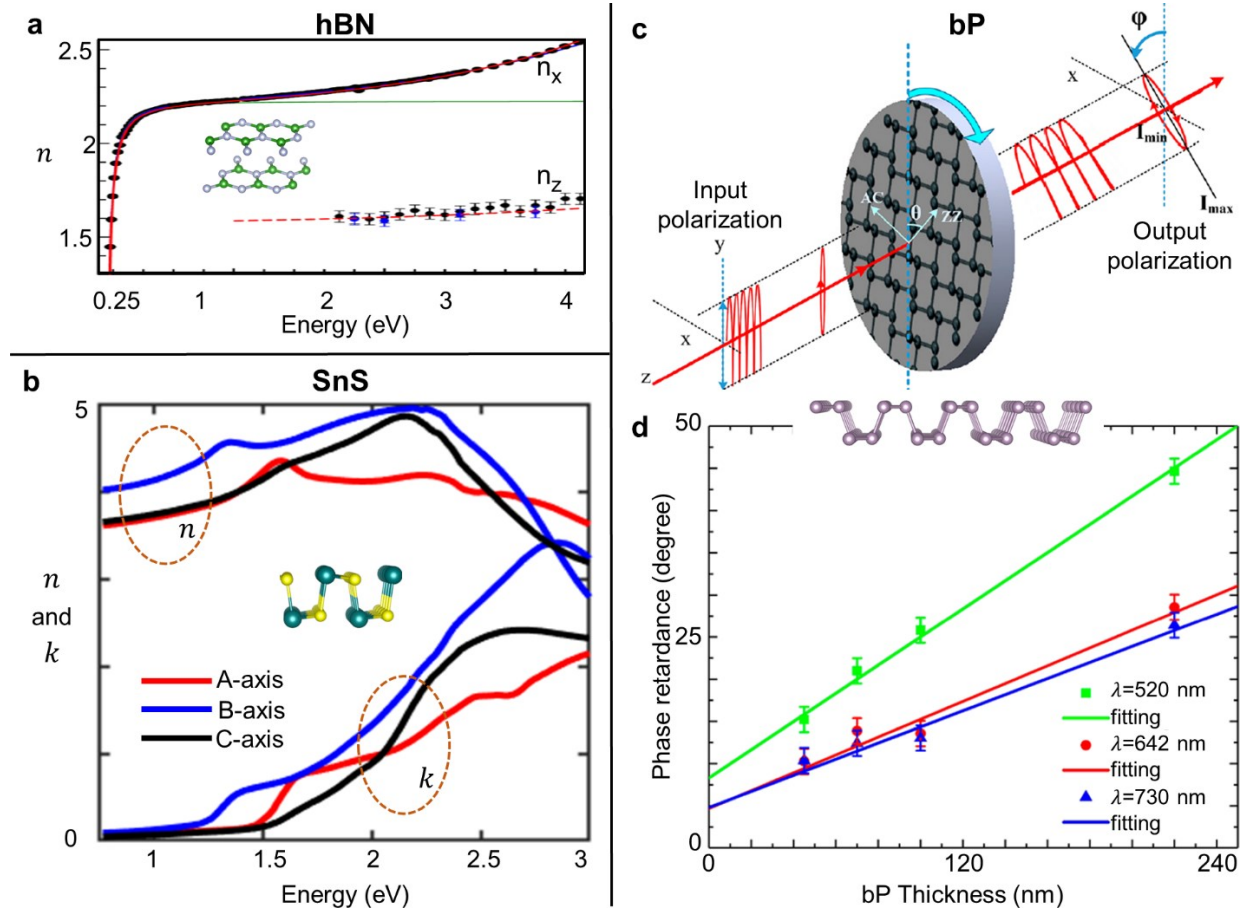

**Figure 7:** Measurement and uses of biaxial and triaxial properties of LMs. (a) Measurement of birefringence of hBN; reprinted with permission from the American Physical Society.[49] (b) Measurement of triaxial SnS.[34] (c-d) Using the in-plane anisotropy of triaxial bP to modulate light polarization; reprinted from *ACS Photonics*.[50] The experimental setup is shown in (c). The accumulated phase is shown in (d) and is proportional to the thickness of bP.

*Electro-optics*

So far we have discussed the linear optical properties of LMs. The linear electro-optic (EO, or Pockels) effect is a non-linear property that can be exploited for electrically-driven optical phase modulators. The change in refractive index due to applied electric field is $\Delta n = -\frac{n^3}{2} r_{hk} E_k$, where $r_{hk}$ are EO coefficients. This $n^3$-dependence of the EO effect, combined with linear-in-$n$ dependence of mode confinement, results in device size that can be scaled down with $n^4$ for a fixed supply voltage.

EO coefficients follow from crystal symmetry, and may be zero for some light propagation directions. For a large number of centro-symmetric LMs (for example, 2H-WS$_2$), the EO effect is not present for even layer numbers, and is small for odd layer numbers greater than unity. For monolayers, the modulation depth (for below-band gap light) is very small due to low light-matter interaction (owing to ultra-small thickness), but can be enhanced by device design such as in the travelling-wave Mach-Zehnder interferometer (MZI) geometry described in section concerning experimental techniques.

The linear EO effect is present in certain LMs with reduced symmetry.[52,53] The theoretically-predicted EO coefficients are fairly low (see **Table 2**), for example $r_{12} = 2.9$ pm/V for GaTe.[52–54] However, LMs could offer useful EO optical phase modulation due to the large refractive index (*e.g.* $n_{GaTe} \sim 3$); for a given applied electric field, $\Delta n_{GaTe}/\Delta n_{LiNbO_3} \sim 1/3$.[53,55] EO effects can be enhanced beyond the intrinsic material properties by breaking symmetry with artificial layer stacking. In recent experimental work, layers of inversion-symmetric 2H-MoSe$_2$ were stacked in AA geometry, resulting in enhanced nonlinear properties.[56] Alloys may feature EO coefficients that are enhanced relative to the pure components, such as MoSSe that is predicted to have EO coefficient five times larger than either MoS$_2$ or MoSe$_2$; EO and nonlinear effects may also be enhanced by Janus phases.[57,58]

| *Material and stacking* | $n_x$ | *Band gap (eV)* | *EO coefficient r (pm/V)* | $r * (n_x)^3$ |
|---|---|---|---|---|
| β-GaS monolayer | 2.7 | 2.5 | 2.4 (theory[53]) | 46 |
| β-GaTe monolayer | 3.1 | 1.6 | 2.9 (theory[53]) | 85 |
| MoS$_2$ monolayer | 3.9 | 1.8 | 1.1 (theory[54]) | 66 |
| MoSSe | 4.1 | 1.6 | 6.7 (theory[57]) | 454 |
| WS$_2$ | 3.8 | 1.9 | 1.3 (theory[54]) | 66 |
| LiNbO$_3$ | 2.3 | 4 | 20 (exp[55]) | 243 |

**Table 2:** Calculated EO coefficients, $n$, band gap, and EO figure of merit $r * (n_x)^3$ for LMs.

**Strong optical phase modulation: Phase transformations and ferroelastic domain switching**

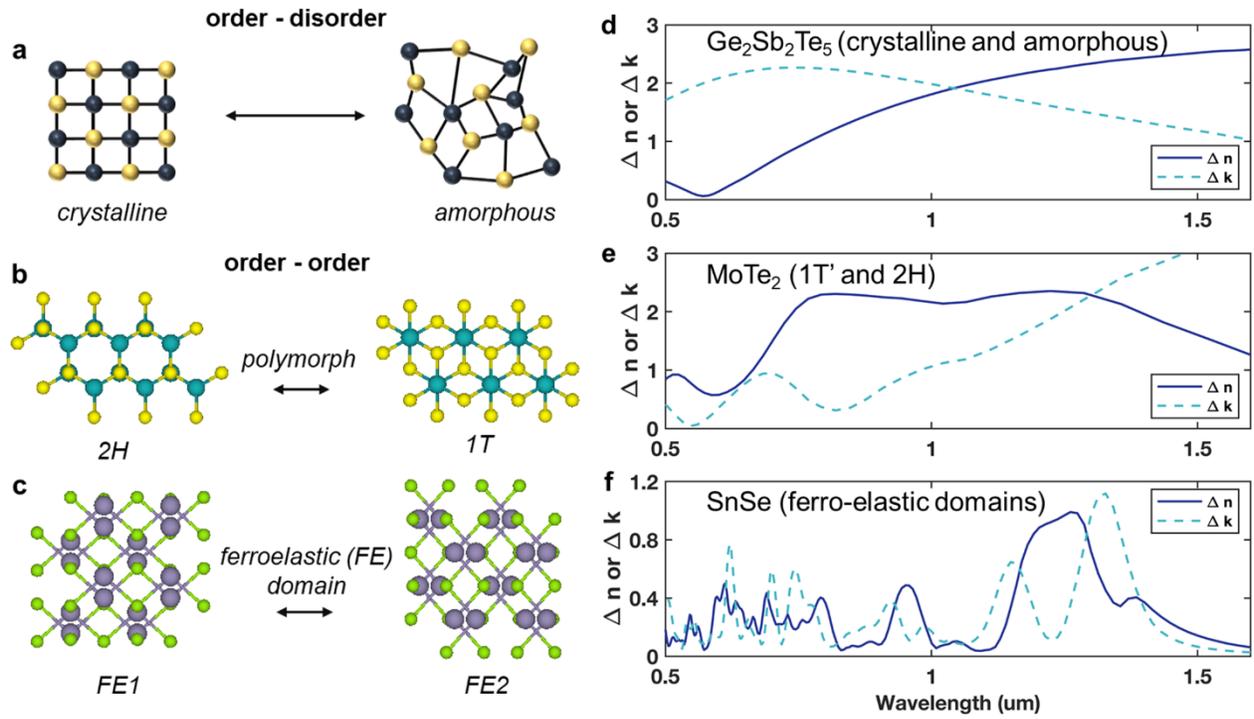

**Figure 8:** (a-c) Schematic illustrations of material transformations enabling strong optical phase modulation. (a) Order (crystalline) – disorder (amorphous) switching, as in the GST system. (b) Martensitic, order-order transformation between TMD polymorphs. (c) Ferroelastic domain switching in GeS-type LMs with low in-plane symmetry. (d-f) Change in real ($n$) and imaginary ($k$) refractive index across functionally-useful material structure transformations. (d) Order-disorder transformation between crystalline and amorphous phases of $Ge_2Sb_2Te_5$.[23] (e) Order-order transformation between 1T' and 2H phases of bulk $MoTe_2$.[29] (f) Ferroelastic domain switching in monolayer SnSe.[15] The structure diagrams were generated using VESTA 3 software.[31]

Mature technologies for optical phase-shifters - such as carrier injection in silicon, or the Pockels effect in $LiNbO_3$ - are based on perturbative changes to the refractive index. In nearly all cases, the refractive index modulation $\Delta n$ is ~ 0.01 or smaller, equivalent to an absolute change of 0.1 - 0.5%.[7–9] In this section we consider materials that can provide strong, non-perturbative optical phase modulation. We first introduce the well-known phase-change materials (PCMs), such as GST, that provide $\Delta n \sim 1$ by virtue of order-disorder transformations. We then consider the outlook for LMs to combine strong optical phase modulation with other desirable attributes including low optical loss and fast, low-energy switching.

PCMs are based on order-disorder transformations (**Fig. 8a,d**) in chalcogenides.[16,23] PCMs have a long history of development for optical and electrical data storage, and processing for device integration is mature.[17] The phase-change functionality is based on time-temperature processes, and can be triggered electrically and optically. PCMs also have drawbacks that may limit their application as active optical phase shifters in photonic integrated circuits. The ordered state is often semi-metallic, and the disordered state is a low band gap semiconductor ($E_g$ ~0.5 - 1 eV). As a result PCMs have substantial optical loss, with $k \gtrsim 1$ in one or both phases throughout the NIR, although clever device design can mitigate the loss.[2] Switching is based on melt-quench and thermal recrystallization processes, which limit operation to sub-GHz frequencies and are inherently energy-intensive.[17] Nanoscale device and materials engineering may extend the frequency range and lower the energy consumption of PCMs.[59] PCMs also suffer from fatigue, which limits device life cycles; even state-of-the-art PCMs for electrical memory are limited to ~$10^9$ switching events, which may be sufficient for memory applications (including photonic neural networks), but is grossly insufficient for telecommunications.[17]

Although the term "phase-change materials" usually refers to chalcogenides such as GST with functionally-useful order-disorder transformations, any material phase transformation may provide strong modulation of optical properties, including order-order transformations in LMs. In materials science, collective atomic displacements that convert between two ordered crystal structures are called martensitic transformations. TMDs feature a panoply of polymorphs (**Figs. 3, 8b**) with contrasting electrical and optical properties, and in recent years there has been sustained interest in using martensitic transformations for device technology.[20,60] These transformations can be triggered by electrical fields, charge injection, and strain - these non-thermal mechanisms are exciting for fast and low-power switching.[21,22,60] The small and collective atomic displacements during switching, facilitated by the layered crystal structure, suggest that material fatigue may be lower and device life cycles may be much longer than for PCMs based on order-disorder transitions.[21] Most of the literature on this topic has focused on transformations between low- and high-conductivity states for electronics and computation.[22] Here we focus instead on proposed applications in photonic integrated devices. In **Fig. 8e** we present the difference in refractive index between the 1T' and 2H phases of $MoTe_2$, which has been shown to support fast, non-thermal switching.[22,29]

The diversity of TMD compositions and phases provides a rich materials design space. For a given transition metal, the energy required for switching between polymorphs tends to decrease with heavier chalcogen, from S, to Se, to Te (**Fig. 9a**).[60] Fast, repeatable, electrically-driven switching at room temperature has been demonstrated in transition metal ditellurides, whereas switching in $MoS_2$ requires elevated temperature to overcome energetic barriers.[16,17] However, for semiconducting phases the band gap also decreases with heavier chalcogen, resulting in substantially higher optical loss for ditellurides than for disulfides and diselenides.[24,29] Furthermore, many TMD phases are metallic or semi-metallic, which also implies high optical loss. In this landscape of material attributes and trade-offs, alloy design may be a powerful strategy to identify materials with optimal combinations of refractive index contrast, optical loss, switching energy, and switching speed for TMD-based optical phase modulators.[24] In **Fig. 9b** we show an example of a theoretically-predicted free energy-composition diagram that guides research on design of transition metal disulfide alloys.[24] Such diagrams indicate compositions that may feature martensitic transformations with vanishingly low switching energy, and advise on materials processing methods. For instance, the diagram shown in **Fig. 9b** predicts that the free energy for switching between 2H and 1T polymorphs will vanish near composition $(MoS_2)_{0.5}(TiS_2)_{0.5}$, and that low-temperature processing methods should be employed to avoid spinodal decomposition.

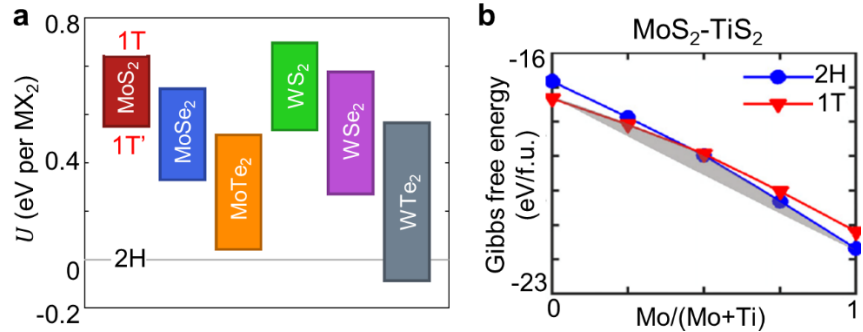

**Figure 9:** The energy required for martensitic transformations varies with material composition; understanding these thermodynamic trends guides materials selection for strong optical phase modulation. **(a)** Calculated energy differences between polymorphs for Mo- and W-based TMD monolayer TMDs; reprinted with permission from Springer Nature.[60] **(b)** Theoretically-predicted Gibbs free energy-composition diagram at 300 K for the $MoS_2$-$TiS_2$ binary alloy system; results are for bulk materials; reprinted with permission from American Institute of Physics.[24]

Electrostatic gating can modulate the optical properties of TMDs even without inducing a phase transformation. These effects have been studied primarily for absorptive/emissive applications, such as light emitting diodes, but recent results show promise for refractive applications.[3,25] Useful field-effect modulation has been demonstrated in $WS_2$ and $MoS_2$ integrated with micro-ring resonator devices operating near $\lambda$ =1570 nm, well below the band gap (**Fig. 10a-b**).[3] In the NIR, far from exciton absorption resonances, the optical loss is reduced and high cavity $Q \approx$ 120,000 can be maintained during switching. Field-effect switching can be enhanced by designing materials that are thermodynamically-adjacent to a polymorphic phase transformation, amplifying the effects of electrostatic gating through non-thermal, collective atomic displacements.

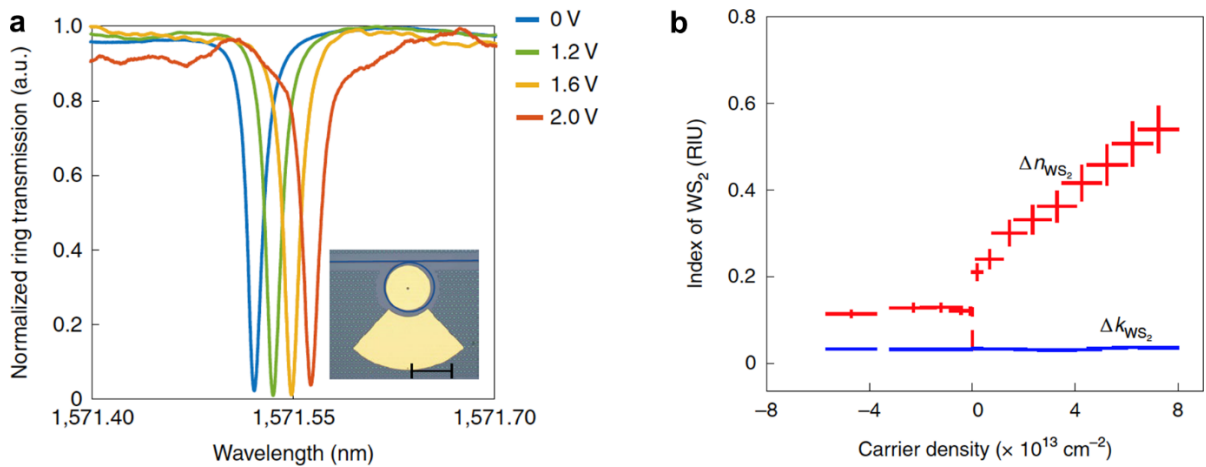

**Figure 10:** Field-effect tuning in the NIR.[3] (a) Tuning the transmission of a micro-ring resonator with gate voltage (line series) applied across a $WS_2$ monolayer partially covering the $Si_3N_4$ waveguide; inset shows an optical micrograph of the device. (b) Change in the complex refractive

index with gate-tuned carrier density in WS$_2$. Images reprinted with permission from Springer Nature.[3]

Materials with substantial optical and dielectric anisotropy also present opportunities for non-thermal switching between domain orientations, rather than between phases. Crystallographic domain orientation can be switched by an applied electric field, and the optical anisotropy can be used for devices such as mode converters or polarizers. A well-known example of non-thermal, electric field-driven structural change is ferroelectric domain switching, which is driven by an energy term linear in applied field. Ferroelastic (FE) domain switching is a newer concept that is based on anisotropy in dielectric susceptibility, and is driven by an energy term that varies quadratically with applied field.[20,61] LMs in the GeS structure-type (**Fig. 3c**), have two FE domain variants related by a 90° rotation. Due to dielectric anisotropy $\varepsilon_{xx} \neq \varepsilon_{yy}$, it is predicted that a sufficiently-strong electric field from a pulsed laser can produce barrier-less conversion between FE domains with substantial optical contrast. In **Fig. 8f** we reproduce the calculated NIR optical contrast between FE domains in monolayer SnSe; the predicted $\Delta n > 1$ and $k < 0.2$ compares favorably to phase-change materials.[15]

**Measuring the optical properties of layered materials**

In this section we review techniques available to measure the optical properties of LMs. Optical property measurements of LMs are frequently complicated by the large out-of-plane birefringence, and the difficulty in preparing crystals with optically-smooth faces with orientation other than the basal plane. Measurements of materials for integrated photonics are also complicated by the differences between reference samples (*e.g.* bulk single crystals), and the active material in photonic integrated circuits (*e.g.* thin films, often with spatially-inhomogeneous grain structure). We summarize the techniques discussed in **Table 3**.

| *Method* | *Experimental Schematic* | *Pros* | *Cons* |
| --- | --- | --- | --- |
| Ellipsometry | 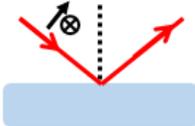 | independent measurement of *n* & *k* | highly-sensitive to surface conditions |
| Normal reflection | 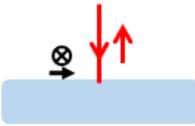 | simple geometry; high spatial resolution | *n* & *k* not independently determined |
| Transmission | 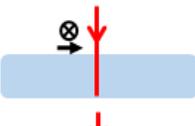 | good for measuring *k* below-band gap | requires flat, parallel surfaces; sample thickness must be known; sample thinning may be required |
| Polarized light transmission | 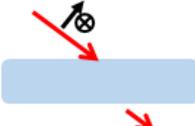 | measure birefringence of LM | same as transmission |

| Waveguide / device integration | 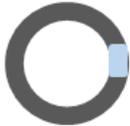 | prediction of performance in intended devices, including processing-dependent effects | results may not reflect intrinsic material properties |
| --- | --- | --- | --- |
| Photo-thermal deflection spectroscopy | 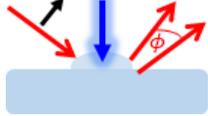 | good for measuring very small $k$ | knowledge of thermal properties needed for data analysis |

**Table 3:** Summary of optical properties measurement techniques. Black arrows indicate the optical electric field polarization, red arrows indicate the direction of light propagation, blue arrow (for photo-thermal deflection spectroscopy) indicates a heating beam.

*Reflection-geometry optical techniques: Ellipsometry, normal-incidence reflectometry, and photothermal deflection spectroscopy*

In ellipsometry we measure the ratio $\rho = r_P/r_S$ of the amplitude reflection Fresnel coefficients.[62] For a simple sample/air interface, this data can be used to directly calculate $n$ and $k$ on a wavelength-by-wavelength basis, without relying on modeling or Kramers–Kronig (KK) transformations. For more complicated structures with thin layers, such as a native oxide, model-based data analysis is required. Since the measurement relies on optical polarization, and uses rather large angles of incidence ($\theta = 50 - 80°$, see **Fig. 11**), samples with large, mirror-like faces are required; the required sample face size is typically at least 1 mm on a side, although with focusing optics smaller faces (down to $\approx 200\ \mu$m) may be used. Ellipsometry is extremely sensitive to surface layers, and can give misleading results if an inaccurate model is used to analyze the data.

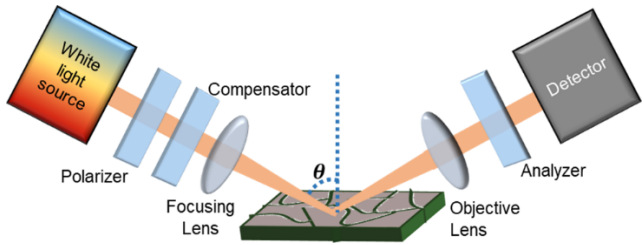

**Figure 11:** Schematic of the ellipsometry technique. The optical elements are, in order: broadband white light source, polarizer, rotating compensator, micro-spot objective, sample, analyzer, and detector.

Normal and near-normal reflectivity measurements are more convenient and more commonly used than ellipsometry. The method can be implemented in a variety of geometries, including confocal microscopy with diffraction-limited spot size under 1 $\mu$m. For normal incidence the Fresnel reflection intensity coefficient is $R = \frac{(n-1)^2 + k^2}{(n+1)^2 + k^2}$; since the experiment yields one measured value ($R$) at each wavelength, modeling and/or KK transformations are required to estimate both $n$ and $k$. Relying on KK transformations can lead to substantial systematic error, especially near the perimeters of the measured data range due to the form of the KK transformation integrand.

Both ellipsometry and reflectivity can be adapted to measure optical anisotropy. For normal-incidence reflectivity, in-plane optical anisotropy can be estimated from polar reflectivity plots using linearly-polarized light. Spectroscopic ellipsometry can in principle measure the full optical response tensor of a material, using an approach called generalized ellipsometry.[62] However, materials with large $n$ (such as the LMs discussed here) strongly refract incident light towards the surface normal, making out-of-plane optical property measurements prohibitively difficult using only exposed basal surfaces.[24] Measurements on polished, non-basal plane surfaces (*i.e.* prismatic faces) have been reported, but the sample preparation is challenging, as it requires an unusually thick crystal, and mechanical polishing risks inadvertent layer folding (imagine grinding and polishing the edge of a stack of paper).[34] As a result, out-of-plane optical property measurements on LMs are rare. Recent work combining ellipsometry and near-field optical measurements may be useful to measure out-of-plane anisotropy.[63,64]

Reflection-geometry methods often present difficulties in measuring the optical properties of materials with low loss, such as semiconductors below the band gap. Coherent reflections from the front and back surfaces of films and crystals - so-called etalon effects - result in intensity oscillations that must be carefully modelled to extract material optical properties.[29] Photo-thermal deflection spectroscopy is a method that excels at measuring optical loss in low-loss materials; this less-common technique may be particularly useful for quantifying optical loss in LMs intended for refractive applications, although it requires understanding of thermal properties that may not be available for all materials.[65]

*Transmission-geometry optical techniques*

Transmission-geometry measurements are well-suited for measuring the optical properties of materials in the low-loss spectral region, such as accurate measurement of $k$ below-band gap, and quantification of defect-assisted absorption. Transmission-geometry measurements of LMs can be found in the older literature, but seem to have gone out of favor.[32,38,39] Measurement of interference fringes in the transmission of highly-convergent, polarized light can quantify birefringence $\Delta n$ in LMs; this method has been been used to determine $\Delta n \sim 1$ for $MoS_2$ between 1 - 1.5 $\mu$m.[38] Recent work used multiple angle of incidence transmission technique for measuring large out-of-plane anisotropy in hBN (also shown in **Fig. 7a**), suggesting use of these LMs in anisotropic components.[49]

The challenges of thinning and polishing samples may inhibit the widespread use of transmission-geometry optical techniques for LMs. Another challenge is inaccurate measurements of sample thickness, which can result in inaccurate estimates of $k$. Thickness can be estimated by counting interference peaks, but this method works only for samples thicker than $\sim 0.1$ $\mu$m; for thinner samples, it is difficult to estimate thickness.[38] Advances in the synthesis of LM films, including large-area, single crystal 2D monolayers, may warrant a revival of transmission-geometry optical techniques.

*Using integrated photonic devices for materials characterization*

The reflection and transmission methods described above are typically intended to measure the "bulk" optical properties of a material; even for thin films, these experiments and their interpretation report thickness-averaged properties. However, optical fields in photonic integrated circuits are highly non-uniform, and the optical properties - and particularly the loss - of a given material in a photonic integrated device may depend strongly on the exact geometry of the device, the material microstructure, and the presence of interface regions. These are processing- and geometry-dependent details that can-not be anticipated by bulk optical property measurements. We therefore emphasize the role of using integrated photonic devices for material characterization.[66,67] In **Fig. 12a** we show an example of the optical loss in $MoS_2$ measured by integration in a micro-ring resonator.[67] By measuring transmission spectrum before and after $MoS_2$ transfer on the ring-resonator (**Fig. 12b**), and calculating change in quality factors ($Q$), the optical loss can be determined. Measurements of cavity properties are combined with modeling of the mode profile and overlap to estimate optical loss. The optical loss measured in such a resonator (at 633 nm) is 850 dB/cm compared to simulated loss of 1390 dB/cm. The difference between measured and simulated loss suggests that device based measurements are important to complement reference crystal measurements.

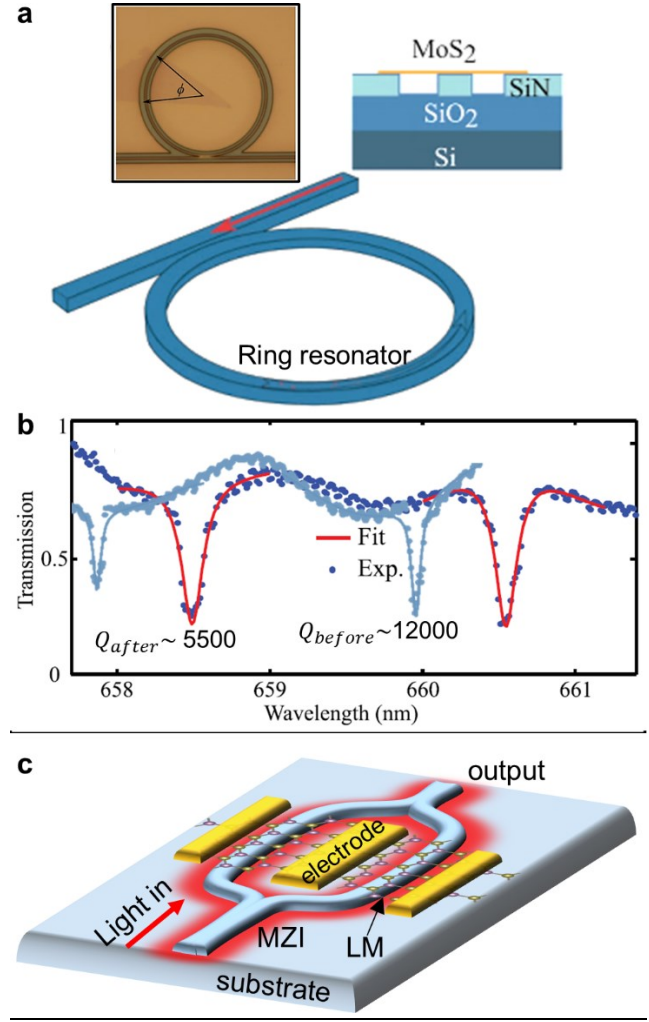

**Figure 12**: Using photonic integrated circuits to measure the optical properties of LMs and guide materials selection. (a, b) Measuring optical loss in $MoS_2$ using a ring resonator; reprinted with permission from American Institute of Physics.[67] (a) Device schematics and optical micrograph showing a $MoS_2$ flake subtending angle ϕ. (b) Transmission spectrum of the resonator before (cyan) and after (blue-red) $MoS_2$ coverage. Quality factors ($Q$) are extracted from Lorentzian fits to the resonances. (c) Schematic of Mach-Zehnder interferometer (MZI) device for measuring EO coefficients of LMs.

*Electro-optics*

For measuring the EO coefficients, a crystal can be processed in a Pockels cell geometry. By modifying the positioning of the crystal with respect to the optical field and applied electric field in the Pockels cell, the EO tensor can be measured. For on-chip integrated photonic-device measurements, the traveling wave MZI geometry may be used, where the electric modulation field co-propagates with the signal light field in two separate arms (**Fig. 12c**). The phase difference is equal and opposite over the two arms, and accumulates over the length of the device.[9]

Pockels effect measurements require large bulk crystals, or thin films on transparent substrates. Further, MZI measurements necessarily require device fabrication. Thus methods to measure EO coefficients in an optical, non-device framework are sorely lacking. However, EO coefficients can be indirectly determined from measurements of the second-order nonlinear susceptibility, since the EO and second order processes are related.[68]

**Synthesis and processing**

Using LMs in photonic integrated circuits will require reliable methods to make and pattern uniform, wafer-scale thin films. Wafer-scale photonic integrated circuitry will be built on established material platforms, including Si, $Si_3N_4$, and InP, using planar processing methods, and it is important to be able to insert LMs into these process flows. In this section we review synthesis methods for LMs, with emphasis on wafer-scale processing.

Processing temperature is an important consideration for integration. Si photonic integrated circuit wafers with passive waveguides patterned from $SiO_2$ or $Si_3N_4$ can withstand at least 800 °C without degradation, which is sufficient for most LM film processing methods. If CMOS devices (such as doped modulators or detectors) are present on the wafer, then the processing temperature should be lower than 600 °C; this may be a challenge for many LMs. Layer transfer methods can decouple the LM film processing temperature from the photonic integrated circuit wafer thermal history.

Thin-film synthesis methods can be grouped according to the phase relation at the surface of the forming film. Here we focus on vapor-phase methods, for which the desired film grows at a vapor-solid interface. Upon close inspection, some vapor-phase methods may be related to liquid- or solid-phase growth, such as the case of conversion of a metal film to a chalcogenide LM. Such distinctions hold interest for materials science, but are not discussed here. Liquid-phase methods capable of coating large areas at relatively low cost are intriguing, but also are not discussed here.

*Solid-source chemical vapor deposition (ssCVD)*: In this process, solid precursors vaporize and react on a substrate to grow a film. This is the most common method for making LM films - the archetypical process is growth of $MoS_2$ films from powder $MoO_3$ and S sources - and is usually referred to simply as CVD.[69] However, it is useful to distinguish between gas- and solid-source CVD. Traditionally, CVD implies that sources are delivered to the reactor as gases, which holds important benefits for wafer-scale and reliable deposition. The use of solid sources means that ssCVD is more difficult to scale.[19]

A great many LMs can be synthesized by ssCVD. The list includes TMDs with transition metals Ti, Zr, Hf, V, Nb, Ta, Mo, W, Re, Pt, Pd, and Fe, and chalcogens S, Se, and Te;[70] main-group metal chalcogenides GeS and GeSe;[71,72] bP.[73]

*Gas-source CVD and metal-organic CVD (MOCVD)*: In gas-source CVD and its variant, metal-organic CVD (MOCVD), gas-phase precursors react on a substrate to grow a film. Of all the LM thin film synthesis methods demonstrated to-date, MOCVD offers the best combination of process uniformity and scale, layer number control, and material quality.[66,67] CVD and MOCVD are versatile, but the requirement of gas-source precursors has limited the range of materials made to date, compared to ssCVD. Materials made by CVD and MOCVD include $MoS_2$, $WS_2$, $WSe_2$, $TiS_2$, SnS, and SnSe.[74–78]

A variation on CVD and ssCVD is to use a chalcogen source (*e.g.* $H_2S$ or solid S) to convert a metal thin film to a LM compound thin film.[79–81] These are two-step processes, first involving deposition of the precursor thin film, often by evaporation or sputtering, followed by conversion in a CVD reactor. These two-step methods are scalable and reliable, but offer inferior layer number control and material quality than ssCVD, CVD, or MOCVD.

*Atomic layer deposition (ALD)*: ALD is a variation on MOCVD in which the film growth is kinetically limited by a precursor delivery sequence. ALD growth of LMs $MoS_2$, $WS_2$, $MoSe_2$, $WSe_2$, $ZrS_2$, $HfS_2$, $TiS_2$, SnS, SnSe, GeS, and GeSe have been reported [82–89]. ALD offers excellent control over film morphology and layer number. However, the kinetic control inherent to ALD dictates rather low reaction temperature, and high-temperature post-deposition annealing is often required to improve film material quality.

*Physical vapor deposition (PVD)*: PVD covers a number of vacuum deposition methods including thermal evaporation, electron-beam evaporation, sputtering, and pulsed laser deposition (PLD). Chalcogenide films deposited by PVD are often anion-deficient, due to the differential re-evaporation of metals and chalcogens. To compensate, an additional chalcogen source may be introduced during film growth, resulting in reactive PVD. PVD is less-widely used for LM film synthesis than ssCVD, CVD, and MOCVD. A number of main-group metal monochalcogenides that evaporate congruently and that have fairly high vapor pressure have been synthesized as thin films by PVD, including SnS, SnSe, GeS and GeSe.[90–93] bP and TMD thin films have been synthesized by PLD and sputtering.[94–96]

*Molecular beam epitaxy (MBE)*: MBE is a variant of PVD that emphasizes epitaxy and precise control over film thickness and morphology. MBE growth of many LMs is challenged by the vastly-different thermodynamic properties of transition metals and chalcogens.[97] Taking for instance $MoS_2$, the high melting point of Mo (2623 °C) means that ad-atom mobility is negligible at substrate temperatures of 800 °C or below. At the same time, the saturation vapor pressure of condensed sulfur at a substrate temperature of 800 °C is well over 1 atm; this implies that the time during which sulfur resides on the heated substrate and may react with the metal is vanishingly short. Despite these challenges, there has been impressive progress such as the growth of monolayer and high-quality $MoSe_2$ on hBN, and the growth of multilayer $WSe_2$ with grain size ≈ 1 μm.[98,99] Taking a cue from the history of complex oxide MBE, and lessons learned from MOCVD growth of LMs, it seems promising to revisit MBE growth of LMs using metal-organic precursors.[100,101]

| Synthesis Technique<br>Photonics relevant metric | ssCVD | CVD/MOCVD | ALD | PVD | MBE |
|---|---|---|---|---|---|
| **Scalability** | Low | High | High | High | High |
| **Range of materials** | High | Medium | Low | High | Low |
| **Crystal quality and chemical purity** | Medium | High | Low | Medium | High |
| Film layer number control/smoothness | Medium | High | High | Medium | High |
| Conformal coating | Medium | Medium | High | Medium | Low |
| Cost (equipment and operation) | Low | Medium | High | Low | High |

**Table 4:** Summary of synthesis methods and trade-offs available for making LM thin films for use in photonic integrated circuits.

*Layer transfer*

LM monolayer or few-layer sheets can be mechanically exfoliated, from thicker crystals or growth substrates, and transferred to different substrates. Layer transfer can be used to make heterostructures, and allows LM synthesis to be decoupled from device integration: LMs can be synthesized at high temperature and in chemically-aggressive conditions, before being transferred to a wafer in mild conditions.[67,102] This approach holds great promise for expanding the materials set available for photonic integrated circuits. However, wafer-scale synthesis and layer transfer has only been demonstrated for a small set of LMs.[102] The methods need to be optimized for adhesion, interface cleanliness, and selection of sacrificial layers, and the processing-properties relationships relevant for photonics remain to be explored.

*Air Stability*

Air stability is a key consideration for any material to be deployed widely in technology. Uncontrolled oxidation can ruin a device, while controlled and self-limited oxidation can imbue functional properties such as surface passivation.[103,104] For chalcogenides of a particular metal, susceptibility to oxidation by oxygen increases as the chalcogen element varies from S, to Se, to Te.[103,105] For TMDs, air stability also varies with the transition metal.[106] Edges and atomic defects accelerate oxidation by providing thermodynamic local minima for oxygen chemisorption, and by reducing the kinetic energy barriers to oxide growth.[107–109] bP is likewise unstable and degrades rapidly upon air exposure.[110] Monolayer and few-layer films are especially susceptible to oxidation due to extreme surface area-to-volume ratios. For instance, monolayer $WTe_2$ degrades even during the exfoliation process and characteristic Raman peaks disappear minutes after exfoliation.[111] The correlation between crystal defects and oxidation complicate the interpretation of such experiments; controlled studies of oxidation of LM films with varying thickness, morphology and defect concentrations would be most useful.

| *Synthesis Technique* | *Main group (group 14) monochalcogenides* | *Group 6 TMDs* | *Group 4 TMDs* | *bP* |
|---|---|---|---|---|
| ssCVD | $GeS$[71], $GeSe$[72] | $MX_2$ (M = Mo, W, X = S, Se, Te)[70] | $MX_2$ (M = Ti, Zr, Hf, X = S, Se, Te)[70] | $bP$[73] |
| CVD/MOCVD | $SnS$[75], $SnSe$[76] | $MoS_2$ and $WS_2$[74], $WSe_2$[77] | $TiS_2$[78] | |

| ALD | SnS and GeS[83], SnSe[84], GeSe[85] | $MoS_2$ and $WSe_2$[86], $MoSe_2$[82], $WS_2$[87] | $TiS_2$[88], $ZrS_2$ and $HfS_2$[89] | |
|---|---|---|---|---|
| PVD | SnS[93], SnSe[90], GeS[92], GeSe[91] | $MoS_2$[95] | $ZrS_2$[96] | bP[94] |
| MBE | SnS[112] | $MoSe_2$[99], $WSe_2$[98] | $TiSe_2$[113], $ZrSe_2$[114], $HfSe_2$[115] | bP[116] |

**Table 5:** Representative publications reporting synthesis of LM films.

**Summary and outlook**

The anisotropic crystal structures of LMs lead to opto-electronic properties that may be especially useful for photonic integrated circuits. Here we have highlighted refractive properties of LMs, and proposed uses of LMs as optical phase modulators and switches. Developing these applications requires a fresh emphasis on knowing and controlling the complex optical properties of LMs, particularly in the below-band gap, low-loss spectral regions. This will require focused efforts in materials characterization and processing, building on but somewhat distinct from the tremendous recent advances in LM research. The processing-properties relationships relevant to refractive uses of LMs in photonic integrated devices remain largely unknown, and are likely to be distinct from those most relevant to uses in other domains such as microelectronics. For instance, how do grain boundaries contribute to optical loss, and how do they affect the performance of strong optical phase modulators based on material phase transformations? Such questions frame an exciting future for research on LMs for integrated photonics.

We've reviewed techniques for measuring the optical properties of LMs, and highlighted the challenges of measuring the complex refractive index below-band gap in highly-anisotropic semiconductors. Certain methods developed decades ago may be primed for revival, motivated by new applications, and enabled by advances in film processing; examples include transmission ellipsometry and photo-thermal deflection spectroscopy, useful to quantify optical loss due to in-gap defects. We encourage the measurement of in- and out-of-plane optical anisotropy (*i.e.* biaxial, triaxial), and the development of new device concepts taking advantage of these sizable effects in LMs. We also encourage increased research on electro-optic properties of LMs and ensuing applications.

Using LMs in photonic integrated circuits will be enabled by advances in materials processing. Looking ahead to photonic circuit integration, we've reviewed processing methods with emphasis on wafer-scale film synthesis. Recent developments in CVD methods, including MOCVD, are promising for future device integration. Two topics particularly relevant to LMs are layer transfer and air stability. Layer transfer may greatly expand the set of materials available for photonic integrated circuits by decoupling materials synthesis and integration, but will require concerted development for each LM, film synthesis method, and integration process flow. Air stability remains a critical issue for many LMs: few LMs are air-insensitive, some form self-limiting native oxides in air, and some suffer from unimpeded oxidation. The field would benefit from continued, focused study of oxidation processes, and on the functional properties of native oxides (or even chemisorbed oxygen molecules).

The diversity of LM polymorphs with distinct opto-electronic properties motivates research on functional material phase transformations. LMs present opportunities to surpass the performance of established phase-change materials, by using order-order transformations that can be switched by non-thermal stimuli, rather than those based on thermal switching between ordered and disordered phases. Here we've emphasized order-order (*i.e.* martensitic) transformations for strong optical phase modulation. The requirements of strong optical phase modulation, fast and low-power switching, low material fatigue and long device lifetime, low insertion loss, and process compatibility add up to a daunting set of material specifications; we suggest that alloy design may be an effective strategy to balance these demands.

The use of LMs for refractive applications in photonic integrated circuits is at a crossroads. The material properties are certainly intriguing, and key device concepts have been demonstrated. However, it's still far from clear that LMs will be better than established options for strong optical phase modulation, all things considered. The opportunities for collaborations between the materials science and photonics engineering communities are focused and exciting.

**Acknowledgements**

This work was supported by an Office of Naval Research MURI through grant #N00014-17-1-2661. This work was supported by Indian Institute of Science start-up grant.